\documentclass[english]{article}
\usepackage[T1]{fontenc}
\usepackage[latin9]{inputenc}
\usepackage[letterpaper]{geometry}
\usepackage{amstext}

\makeatletter
\def\@biblabel#1{\hspace*{-\labelsep}}
\makeatother
\newcommand{\lyxaddress}[1]{
\par {\raggedright #1
\vspace{1.4em}
\noindent\par}
}

\makeatother

\usepackage{graphicx} 
\usepackage{subfig} 
\usepackage{caption}
\usepackage[titles,subfigure]{tocloft} 
\usepackage{babel}

\begin{document}
%
%
%
%
%


\title{Probabilistic generation of random networks taking into account information on motifs occurrence}

\author{Fr\'ed\'eric Y. $Bois^{\text{}(a,b)}$, 
Ghislaine $Gayraud^{\text{}(c,d)}$}

\maketitle

\lyxaddress{a. Chair of Mathematical Modeling for Systems Toxicology, Universit\'e de Technologie de Compi\`egne, Compi\`egne, France. Email: frederic.bois@utc.fr\\
b. INERIS, DRC/VIVA/METO, Verneuil en Halatte, France.\\
c. LMAC, Universit\'e de Technologie de Compi\`egne, Compi\'egne, France.\\
d. LS, CREST-INSEE, 3, avenue Pierre Larousse, Malakoff, France}



\section*{Abstract}

Because of the huge number of graphs possible even with a small number of nodes, inference on network structure is known to be a challenging problem. Generating large random directed graphs with prescribed probabilities of occurrences of some meaningful patterns (motifs) is also difficult. We show how to generate such random graphs according to a formal probabilistic representation, using fast Markov chain Monte Carlo methods to sample them. As an illustration, we generate realistic graphs with several hundred nodes mimicking a gene transcription interaction network in \textit{Escherichia coli}.


\section*{Introduction}

Graph models are essential tools for understanding and modeling complex systems of interacting variables or agents (Albert and Barab\'{a}si, 2002).  
The global features of social, telecommunication or biological networks can only be analyzed with correspondingly large graph models. Such models consist of bonds (edges) indicating relationships between agents (nodes), with parameters quantifying the strength of the bonds. Over the years, graph models have been extensively studied both in theory (see among others Lauritzen,  1996) and in methodology (Whittaker, 1990)  in different scientific areas, such as statistical physics (Gibbs, 1902), genetics (Wright, 1921), economics (Wold, 1954) or social sciences (Blalock, 1971). When there is no ambiguity about the links between nodes and their interactions strengths, complex systems are well described by deterministic networks. In the presence of ambiguity about the system, probabilistic graphs are usually considered. In that case, bonds represent stochastic links between nodes and their parameters specify conditional distributions. Probabilistic graphs offer a natural framework for statistical inference and knowledge integration. For example, in communication engineering, we may want to know what conditions the robustness of a network; in systems biology we may be interested in understanding how genes control each others.

The theory of random graphs was initiated by Erd\"{o}s and R\'enyi in a series of papers Erd\"{o}s  and  R\'enyi, 1959, 1960, 1961) in which they proposed to generate a random graph with a given number of labeled nodes by connecting every pair of nodes with probability $p$ ($p \in (0,1)$). The main goal of random graph theory has been to determine at which connection probability $p$ a particular property of a graph will most likely emerge. One of the first properties studied by Erd\"{o}s and R\'enyi is the appearance of given sub-graphs like cliques, triangles, \textit{etc.} In some sense, the question they have addressed is relative to the structure/topology on a graph. In many settings, interest focuses first on the structure of a graph and the question arose of whether the random graphs proposed by Erd\"{o}s and R\'enyi were able to display specific structures of real complex networks. Over the past few years, there has been a growing interest in investigating and developing new tools and measures able to capture specific properties of real networks. Among such properties are the degree distribution, corresponding to the probability $P(k)$ that a node in the network is connected with $k$ other nodes (Albert \textit{et al}, 2000;   Leskovec \textit{et al},  2010), small-world properties (Watts and Strogatz, 1998) in which most nodes are not neighbours of one another, or the node clustering coefficient  (Watts and Strogatz, 1998).
 
Nevertheless, inference about network structures remains difficult because of the extremely large number of possible graphs, even with a modest number of nodes (Markowetz and  Spang, 2007). Estimating the amount of data needed to recover the structure of a graph is also difficult, but it is clear at least in biology, that most of the current experimental designs are insufficiently powerful for that aim. In such a context, every bit of information counts. Still in biology, the relevant scientific literature indicates that all graphs are not equally plausible, some being \textit{a priori} more likely than others. Accounting for prior knowledge is well formalized in Bayesian statistics (Robert, 2001), but the probabilistic representation of such knowledge is still a question. Mukherjee and Speed (2008) have recently proposed a set of informative priors for network structure inference. More precisely, they have considered priors able to capture information relative to existence of edges, degree distribution or sparsity structure in Bayesian networks,\textit{ i.e.}, acyclic directed graph models.  

Network motifs are patterns (sub-graphs) that recur within a network much more often than expected for random graphs (Milo \textit{et al}, 2002). It has been shown that gene transcription regulation networks, for example in the bacteria \textit{Escherichia coli}, contain a small set of network motifs (see Alon,  2007 and references therein), suggesting that such motifs are basic building blocks of transcription networks. An important aspect of network topology inference is therefore to include the probability of occurrence of such network motifs (see Janson \textit{et al} (2000) for an overview).

In this article, we address precisely that question. We extend the approach of Mukherjee and Speed (2008)  by relaxing the acyclicity requirement which characterizes Bayesian networks and propose rigorous probabilistic representations of \textit{a priori} information on pairwise links, degree node distribution, and network motifs. 
We use Markov chain Monte Carlo (MCMC) simulations to sample networks satisfying those joint distributions, for moderately large networks (several hundred to thousands of nodes). Such random networks can be used as priors for formal inference, after updating with data in a Bayesian framework. They can also be used for pure simulation purposes, \textit{e.g.} for methods or software testing. We do not deal with data and associated likelihood (or "score") functions, but focus on probabilities. Our distributions, however, are entirely compatible with any score function and can be used for inference, in particular in a Bayesian framework.  As an illustration of our results, we generate realistic graphs mimicking a gene transcription interaction network in \textit{E. coli}. The weight of the various proposed priors is examined.


\section*{Results}

\subsection*{Graph Models for Networks} 

A graph model simply consists of nodes (vertices) connected by edges (Wilson, 2012)  The nodes often represent physical entities (people, genes, servers...) and the edges represent links or dependencies between them ("is a friend", "controls", "is physically connected to", ...). Nodes can be assigned attributes (\textit{e.g.}, "on", "off") which can depend in turn on the attributes of the nodes to which they are connected. The edges may also have attributes influencing those of the nodes they connect. Edges may be "undirected" or "directed", the latter case (often noted by an arrow) introducing an asymmetry between the two nodes. For example, an arrow from node $i$ to node $j$ may indicate that $i$ controls $j$, the reverse being not true. Directed edges can in turn be signed, indicating a positive or negative control, \textit{etc}. Finally, graph models may have global features imposed to them. For example, we may impose no unconnected node. A commonly imposed feature is "acyclicity". In that case, the graph model cannot contain any path (succession of edges) linking any node to itself, and in particular no "auto-loop" edge from a node to itself. In the case of directed edges, paths are understood to follow the directions of the edges. A particular class of such acyclic directed graph models, Bayesian networks, has a clear probabilistic interpretation and is easily amenable to inference about network structure or parameters (Neapolitan, 2003) 
 
Since we are interested in generating general graphs, in particular those describing genetic regulatory systems, we consider directed graph which may be cyclic (see De Jong (2002) for an overview of genetic regulatory networks modeling). More precisely, our graphs are composed of labeled nodes with directed edges. There can be two reverse edges between any two nodes and auto-loops are allowed.

\subsection*{Informative priors on networks}

Let $G$ be a graph described by a set $V(G)=\{v_1,\ldots,v_n\}$ of $n$ vertices ($n \geq 2$) and a set 
$E(G)=\{e_{i,j}: (i,j)\in \{1,\ldots,n\}\times \{1,\ldots,n\}\}$ of directed edges and auto-loops. $G$ may be described by its adjacency matrix $A$, a $(n\times n)$-matrix whose generic term is given by $a_{i,j}=\left\{\begin{array}{ll}1 & \mbox{\rm{if the edge $e_{i,j}$ exists}} \\
0& \mbox{\rm{otherwise}} \end{array} \right.$.  
 
Incomplete \textit{a priori} knowledge on such a graph can be described by a statistical distribution. Given $n$, the number of nodes of $G$, we propose to include three levels complexity in that distribution. It is only mandatory to define the first level, which does not reflect any specific structure except for the probability of presence of individual edges. 
We then refine it by including information on the degree distribution.
The next step is to incorporate information related to the occurrences of sub-network motifs.

\subsubsection*{Priors on individual edges} 

As in the random graphs considered by Erd\"os and R\'enyi, prior knowledge on each individual edge can be conveniently
modeled by a Bernoulli distribution, assigning probability $p_{i,j}$ to the existence of a directed edge from node $i$ to 
node $j$, that is, $e_{i,j} \sim B(p_{i,j})$ for all $(i,j) \in \{1,\ldots,n\}\times\{1,\ldots,n\}$. In this context, the adjacency matrix $A$ related to a graph $G$ with $n$ nodes, becomes $A=(e_{i,j})_{\{1\leq i,j \leq n\}}$.

For a graph of a given size $n$, there are $n^2$ individual pairwise possible links in the cyclic case and at most $n^2 - n$ in the acyclic case (auto-loops being ruled out by definition). Specifying the complete set of edges priors requires the definition of 
a $n \times n$ matrix $\mathbf{P}=(p_{i,j})_{1 \leq i,j \leq n}$ (with a null diagonal in the acyclic case). If the pairwise links are supposed to be independent, the probability distribution for the entire graph $G$ is therefore

\begin{eqnarray}
  P_{Bern,G} = \prod_{i,j=1}^{n} p_{i,j}^{e_{i,j}}(1 - p_{i,j})^{e_{i,j} - 1}.   \label{Bernoulli-Proba}
\end{eqnarray}

There are various ways to choose or elicit values for the individual prior probabilities $p_{i,j}$, which we will further discuss in our application 
but intuitively they are related to the weight of the prior evidence ({\it e.g.}, {\it p}-values (Bernard and Hartemink, 2005) we have on the existence of given edges. Other distributions could be used, such as a =multinomial if we had chosen to give a sign 
to the edges (to indicate positive or negative interactions, for example). But the principle would remain the same, and in the 
absence of precise prior information, a Bernoulli prior is probably all we can specify.

\subsubsection*{Priors on degrees' counts} 

The degree $deg(v)$ of a vertex $v$ is the total number of edges to which vertex $v$ participates. The degree distribution of a graph $G$ is a function $P(d)$ expressed in terms of $|\{v \in V(G) : deg(v) = d\}|$, the total number of vertices having degree $d$. In many biological networks (Jeong \textit{et al}, 2000),  it appears that the degree distribution has a power-law tail, which means that $P(d)\propto d^{-\gamma}$, with $\gamma >0$. Such networks are called scale-free (Barab\'{a}si and Albert, 1999). Then, we define the probability distribution of a graph $G$ as follows 

\begin{eqnarray}
P_{deg,G} & \propto& P_{Bern,G} \times \prod_{i=1}^{n} \left( \sum_{j \in \{1,\ldots, n\}: \sum_j e_{i,j} >0} e_{i,j}\right)^{-\gamma}. \label{Deg-Proba}
\end{eqnarray}

Since we do not impose that every node should be linked to another one, the degree distribution attributes implicitly a weight of one to any isolated node. It entails that the probability, with respect to degrees, of an empty graph (without any connection between its nodes) is one. 

Here again, other distributions, even empirical (as defined by an histogram) and reflecting better the degree distribution of a given 
class of graphs, could be used if enough information was available.

\subsubsection*{Priors on motifs}

\begin{figure}[!ht] 
\begin{center}
\setlength{\unitlength}{1cm}
\begin{picture}(12,2.5)(0,0) 
\put(0.5,1.5){$\circ$}
\put(1.0,2.0){$\circ$}
\put(1.5,1.5){$\circ$}
\put(0.65,1.65){\vector(1,1){0.4}}
\put(1.54,1.66){\vector(-1,1){0.4}}
\put(1.50,1.60){\vector(-1,0){0.85}}

\put(2.5,1.5){$\circ$}
\put(3.0,2.0){$\circ$}
\put(3.5,1.5){$\circ$}
\put(2.65,1.65){\vector(1,1){0.4}}
\put(3.13,2.04){\vector(1,-1){0.4}}
\put(3.50,1.60){\vector(-1,0){0.85}}

\put(4.5,1.5){$\circ$}
\put(5.0,2.0){$\circ$}
\put(5.5,1.5){$\circ$}
\put(5.54,1.66){\vector(-1,1){0.4}}
\put(5.50,1.60){\vector(-1,0){0.85}}

\put(6.5,1.5){$\circ$}
\put(7.0,2.0){$\circ$}
\put(7.5,1.5){$\circ$}
\put(6.65,1.65){\vector(1,1){0.4}}
\put(7.54,1.66){\vector(-1,1){0.4}}

\put(8.5,1.5){$\circ$}
\put(9.0,2.0){$\circ$}
\put(9.5,1.5){$\circ$}
\put(8.65,1.65){\vector(1,1){0.4}}
\put(9.50,1.60){\vector(-1,0){0.85}}

\put(10.5,1.5){$\circ$}
\put(11.0,2.0){$\circ$}
\put(11.5,1.5){$\circ$}
\put(10.65,1.65){\vector(1,1){0.4}}
\put(11.50,1.60){\vector(-1,0){0.85}}
\put(10.65,1.52){\vector(1,0){0.85}}

\put(0.0,0.5){$\circ$}
\put(0.5,1.0){$\circ$}
\put(1.0,0.5){$\circ$}
\put(0.15,0.65){\vector(1,1){0.4}}
\put(0.65,1.04){\vector(1,-1){0.4}}
\put(1.00,0.60){\vector(-1,0){0.85}}
\put(0.15,0.52){\vector(1,0){0.85}}

\put(1.8,0.5){$\circ$}
\put(2.3,1.0){$\circ$}
\put(2.8,0.5){$\circ$}
\put(1.95,0.65){\vector(1,1){0.4}}
\put(2.84,0.66){\vector(-1,1){0.4}}
\put(2.80,0.60){\vector(-1,0){0.85}}
\put(1.95,0.52){\vector(1,0){0.85}}

\put(3.6,0.5){$\circ$}
\put(4.1,1.0){$\circ$}
\put(4.6,0.5){$\circ$}
\put(3.75,0.65){\vector(1,1){0.4}}
\put(4.64,0.66){\vector(-1,1){0.4}}
\put(4.30,1.10){\vector(1,-1){0.4}}

\put(5.4,0.5){$\circ$}
\put(5.9,1.0){$\circ$}
\put(6.4,0.5){$\circ$}
\put(5.55,0.65){\vector(1,1){0.4}}
\put(6.44,0.66){\vector(-1,1){0.4}}
\put(6.10,1.10){\vector(1,-1){0.4}}
\put(5.89,1.10){\vector(-1,-1){0.4}}

\put(7.2,0.5){$\circ$}
\put(7.7,1.0){$\circ$}
\put(8.2,0.5){$\circ$}
\put(7.35,0.65){\vector(1,1){0.4}}
\put(8.24,0.66){\vector(-1,1){0.4}}
\put(7.90,1.10){\vector(1,-1){0.4}}
\put(7.36,0.60){\vector(1,0){0.85}}

\put(9.0,0.5){$\circ$}
\put(9.5,1.0){$\circ$}
\put(10,0.5){$\circ$}
\put(9.15,0.65){\vector(1,1){0.4}}
\put(9.65,1.04){\vector(1,-1){0.4}}
\put(10.0,0.60){\vector(-1,0){0.85}}
\put(9.15,0.52){\vector(1,0){0.85}}
\put(9.49,1.10){\vector(-1,-1){0.4}}

\put(10.8,0.5){$\circ$}
\put(11.3,1.0){$\circ$}
\put(11.8,0.5){$\circ$}
\put(10.95,0.65){\vector(1,1){0.4}}
\put(11.5,1.10){\vector(1,-1){0.4}}
\put(11.8,0.6){\vector(-1,0){0.85}}
\put(10.95,0.52){\vector(1,0){0.85}}
\put(11.29,1.10){\vector(-1,-1){0.4}}
\put(11.84,0.66){\vector(-1,1){0.4}}

\end{picture}

\caption{The thirteen possible three-node motifs.}
\label{FigureLoops}
\end{center}
\end{figure}
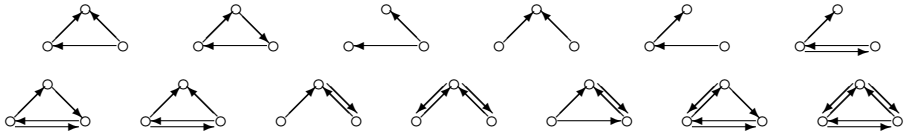

One important local property of networks is the eventual occurrence of motifs. Motifs are defined here following Milo \textit{et al} (2002),  Alon (2007), Shen-Orr  \textit{et al} (2002), Milo   \textit{et al}  (2004) and  Kashtan   \textit{et al}  (2004), as over-represented sub-graphs compared to what is found in an Erd\"os and R\'enyi random graph. Some motifs have a notable importance in biological networks because they can carry out specific information-processing functions, and hence may help in understanding the global behavior of such networks (Masoudi-Nejad, 2012). For example, there are thirteen possible configurations for the relationships between three nodes (see Figure \ref{FigureLoops}). Among the non-degenerate configurations, only the feed-forward loop (FFL, top row, first motif on the left, Figure \ref{FigureLoops}) has been found in the
transcriptional regulation network of \textit{E. coli} (Alon, 2007; Shen-Orr  \textit{et al}, 2002; Mangan \textit{et al},  2003). No feed-back loop (FFB, top row, second motif on the
left, Figure \ref{FigureLoops}) has been observed in \textit{E. coli}. The FFL is one of the most studied network motifs in transcription interactions. It corresponds to a directed sub-graph of three nodes (genes) such that one of them is regulated by the two others, which are linked. Given that each of the regulatory interaction can either an activation or a repression, there are eight sub-types of signed FFL, two of them occurring much more frequently than the other six in transcription networks (Mangan and Alon, 2003; Mangan \textit{et al},  2006). 

In this paper, we consider network motifs with three nodes; auto-loops are not taken into account in such sub-graphs. We define a motif distribution based on the proportion of FBL among all three-node loops. More precisely, for a graph $G$ let us consider $N_1$ 
the number of FBL motifs and $N_2$ the number of FFL motifs and note they may be expressed in terms of 
$(e_{i,j})_{\{ 1 \leq i,j \leq n\}}$ as follows,  $N_1= \displaystyle{\sum_{(i,j,k): i\neq j,k\neq j,i\neq k} 
e_{i,j} e_{j,k}e_{k,i}}$ and $N_2= \displaystyle{\sum_{(i,j,k): i\neq j,k\neq j,i\neq k} 
e_{i,j} e_{j,k}e_{i,k}}$. 

For a graph $G$ with a total number $N_1+N_2$ of motifs of type FBL and FFL, we place a beta-binomial probability with parameters $u$ and $v$ on $N_1$, 
$BB(N_1 | u, v, N_1 + N_2)$. The prior for graph $G$ is then

\begin{eqnarray}
P_{Motif,G} & \propto & P_{Bern,G} \times C_{N_1 +N_2}^{N_1} \frac{B(N_1+u,N_2+v)}{B (u,v)},   
\label{Motif-Proba}
\end{eqnarray}
where $C_{N_1 +N_2}^{N_1}$ is a Binomial coefficient and $B(\cdot,\cdot)$ denotes the Beta function. The choice of a Beta-Binomial distribution is justified by the fact that we may not know the exact proportions of FFL and FBL, but simply have observations about the numbers of such loops in some actual network we base our prior on. Obviously, as we will discuss later, other motifs could be tracked and entered in the definition of a graph probability, using a similar device.

\subsubsection*{Piecing together a global prior}

In addition to the probabilities $P_{Bern,G}$, $P_{Deg,G}$ and $P_{Motif,G}$ defined by (\ref{Bernoulli-Proba}), (\ref{Deg-Proba}) and (\ref{Motif-Proba}) respectively, we consider one more graph distribution $P_{Total,G}$ which combines all informative priors independently. Therefore: 

\begin{eqnarray}
 P_{Total,G} & \propto & \prod_{i,j=1}^{n} p_{i,j}^{e_{i,j}}(1 - p_{i,j})^{e_{i,j} - 1} \times \prod_{i=1}^n \left( \sum_{j \in \{1,\ldots, n\}: \sum_j e_{i,j} >0} e_{i,j}\right)^{-\gamma} \nonumber\\
         &       & \times \quad C_{N_1 +N_2}^{N_1} \frac{B(N_1+u,N_2+v)}{B (u,v)}.    
\label{Total-Proba}
\end{eqnarray}

\subsection*{Application to a Biological Network} 

Transcriptional regulatory networks orchestrate the gene expression of cells. In such networks, the nodes are operons (one or more 
genes transcribed on the same mRNA template). Edges go from operons encoding a transcription factor to operons directly regulated 
by that factor. Shen-Orr {\it et al.} (2002) developed and applied motif-detection algorithms to the transcriptional regulation network of {\it E. coli}. They extracted data from the RegulonDB transcriptional database (Salgado \textit{et al}, 2013), and enhanced them with 
additional transcription factors and interactions described in the literature. We used here the latest version of the dataset (version 
1.1, made publicly available by Dr. U.Alon). 

\begin{figure}[!ht] 
\begin{center}
\includegraphics[scale=0.6]{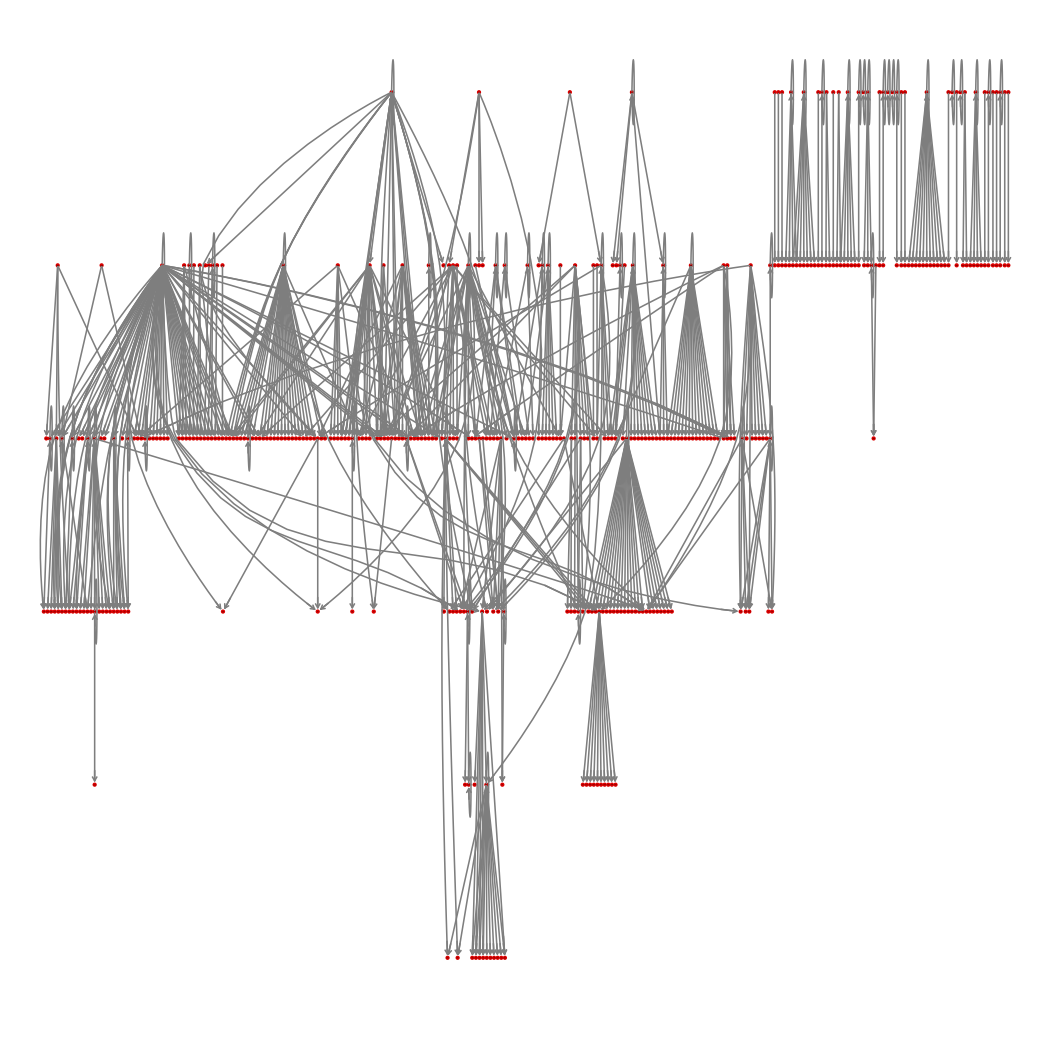}
\caption{{\it E. coli} transcriptional regulatory network, as reported in Shen-Orr  \textit{et al} (2002), with minor updates (see text).}
\label{FigurePriorGraph}
\end{center}
\end{figure}

A graph representation of the {\it E. coli} transcriptional regulatory network is shown on Figure \ref{FigurePriorGraph}. 
It contains 423 nodes, all connected, with 578 directed edges. That is actually only 0.32\% of the number of possible edges, 
indicating that the network is sparse.

We report here the results for Alon's full size network (of 423 nodes). We investigate here which elements of our prior knowledge on {\it E. coli} regulatory network features are the most important to simulate realistic networks.

\begin{figure}[!ht] 
\begin{center}
\includegraphics[scale=0.6]{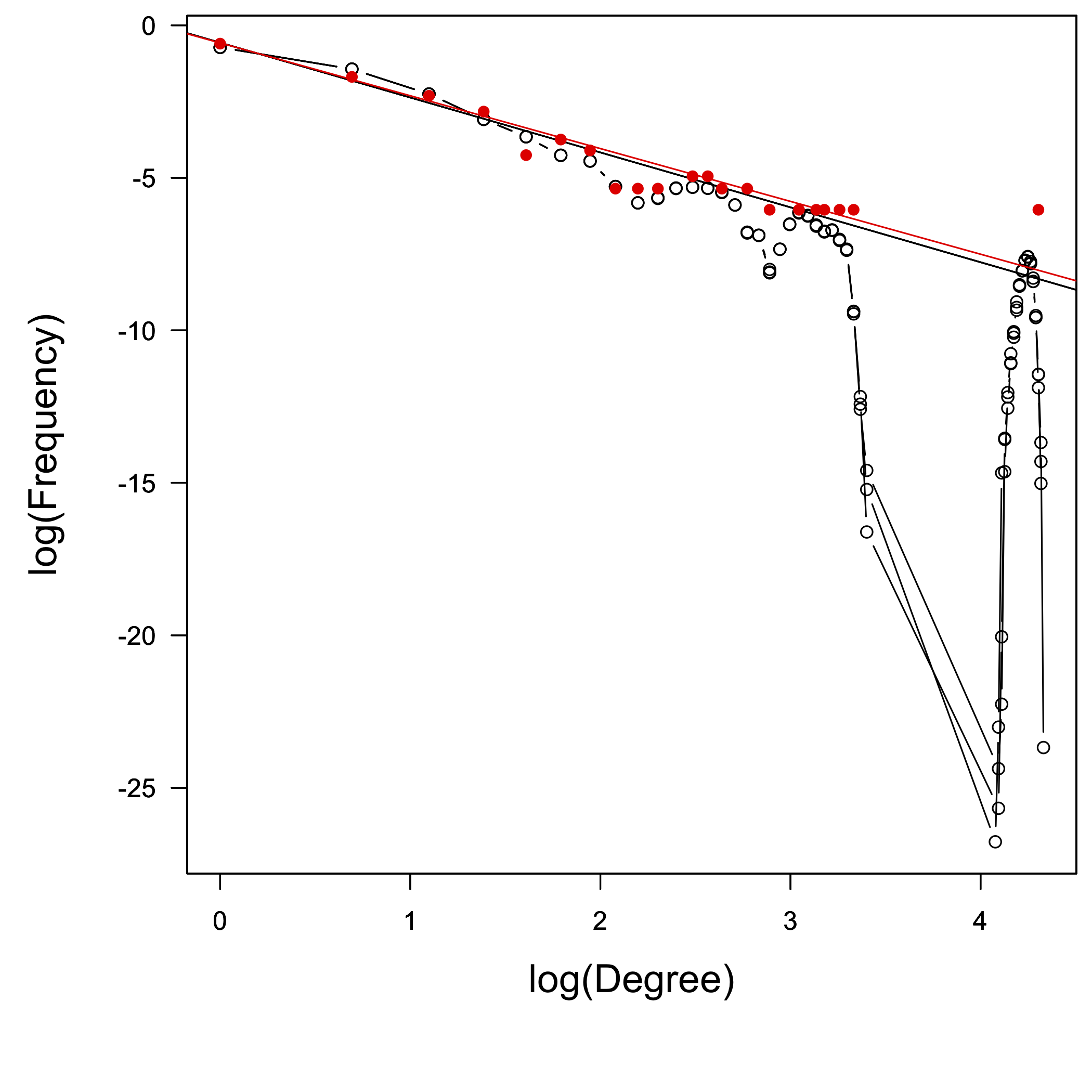} 
\caption{Degree distribution in {\it E. coli} actual transcriptional regulation network (in red) and in Monte-Carlo sampled networks (in black). Informative distribution on individual edges. The dip for high degrees 
due to a deviation of reality from the power law assumption}
\label{FigureDeg}
\end{center}
\end{figure}

In a first set of simulations we assigned "vague" priors to the edge probabilities, setting them all to the same value (equal to $578 / 423^2$, {\it i.e.}, 0.0032). With that prior, all connections are equally probable, and their expected number 
is equal to the one observed for {\it E. coli}. In {\it E. coli} the degree distribution follows approximately a power law
with an exponent of 1.7 (see Figure \ref{FigureDeg}), so we set $\gamma$ to that value when the degree distribution was in effect.
In addition, {\it E. coli} regulatory network is known to contains 42 FFLs and no FBL. To allow flexibility in the prior and allow some occurence of FBL motifs, we set equation (\ref{Motif-Proba}) parameter $u$ to 2 and parameter $v$ to 50 in our simulations. That implies an expected proportion of only two percents FBLs.

In the second set of simulations we use the same priors for degree distribution and motifs frequencies, but used informative priors to individual edge probabilities: The edges reported by Alon {\it et al.}, were assigned probability 0.95. Non-reported, therefore hypothetical, edges were assigned probability 0.00016. Together those probabilities lead again to 578 expected edges.

\begin{figure}[!ht] 
\begin{center}
\includegraphics[scale=0.6]{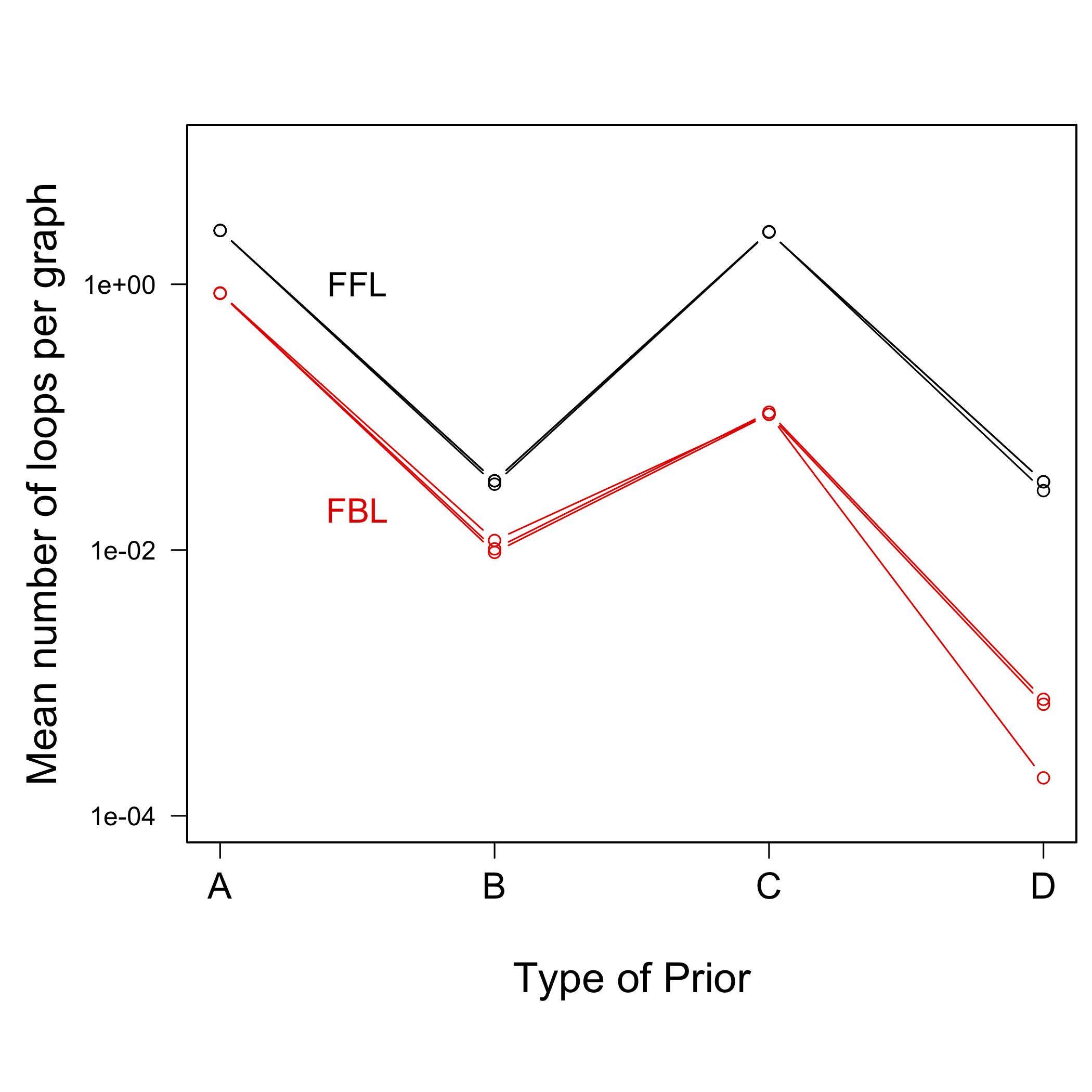} 
\caption{Motifs frequencies in transcriptional regulation networks generated using a vague prior on individual edges. A: prior on individual edges only; B: prior on individual edges and degree distribution; C: prior on individual edges and the proportion of feed-back loops; D: all three priors together.} 
\label{FigureMotifsCounts}
\end{center}
\end{figure}

In all cases, three MCMC chains of 2 billions iterations were run independently. Convergence of the edge probabilities (according to Gelman and Rubin's criterion) was always attained after at most 1 billion iterations. The degree distribution ({\it e.g.}, Figure \ref{FigureDeg}) and motif frequencies ({\it e.g.}, Figure \ref{FigureMotifsCounts}) also converged within that time frame: Results from the three independent chains basically overlap, except in the case of rare events (with frequencies less that one in 10,000), where Monte Carlo sampling uncertainty becomes noticeable. We therefore discarded systematically the first billion simulations and base all the following results on the second billion. We optimized our MCMC sampling C code and simulations are rather fast. Running 2 billions iterations to generate a random graph with 423 nodes takes about 2 minutes on a Intel Core 2 Duo machine clocked at 2.13 GHz. It takes actually little more time to sample graphs with a thousand nodes. Overall, the time it takes to update all the elements of the adjacency matrix is approximately proportional to the number of its elements, and therefore proportional to the square of the number of nodes for the graph considered. In our implementation, memory requirements are simply proportional to the number of nodes and minimal.

\begin{figure} [!ht] 
\begin{center}
\includegraphics[scale=0.6]{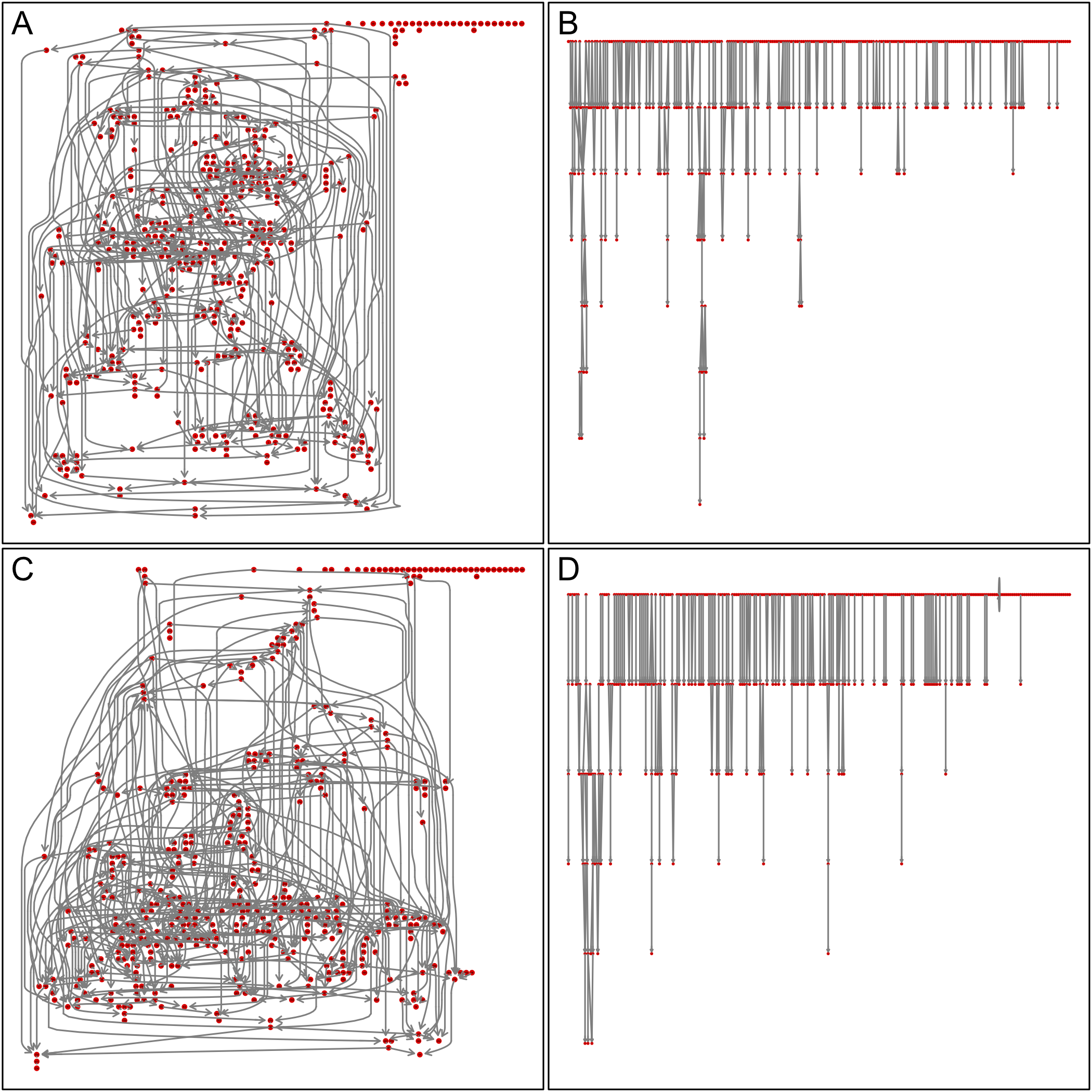} 
\caption{Transcriptional regulation networks generated using a vague prior on individual edges. Red dots: 423 genes in each network. Panel A: prior on individual edges only; B: prior on individual edges and degree distribution; C: prior on individual edges and the proportion of feed-back loops; D: all three priors together.}
\label{FigureFlat}
\end{center}
\end{figure}

Figure \ref{FigureFlat} shows samples of networks generated using the above vague prior on individual edges. Used alone, that prior gives all edges the same probability and the resulting network has little structure, except that the expected number of edges is respected (Figure \ref{FigureFlat}, panel A). The proportion of FBLs is 1/4, as expected in an unconstrained setting (there are two possible FBLs and six possible FFLs for each triplet of nodes). Adding the prior component on degree distribution imposes a major change in network shape. The number of edges is similar, but the structure becomes hierarchical (panel B). Placing a prior on the proportion of FBLs and FFLs, in addition to the prior on individual edges, has little visible impact on the structure (panel C), but the proportion of loops is now much lower and close to its expected value. Finally (panel D) putting the three priors together gives us again a hierarchical structure, but with the correct proportion of FBLs.

Figure \ref{FigureMotifsCounts} shows in more details how the number of feedback and feed-forward loops is influenced by the specification of prior knowledge in the context of a vague specification of individual edge probabilities. The hierarchical structure imposed by the degree distribution (prior "B") leads to a much lower number of loops, but without altering the ratio of FBLs to FFLs (25\% in the case of prior "A" and "B"). In contrast, imposing a prior on the proportion of FBLs can later reduce both the ratio (4\% in the case of prior "C" and 2\% with prior "D") and the number of loops in the network.
 
\begin{figure}[!ht]
\begin{center}
\includegraphics[scale=0.6]{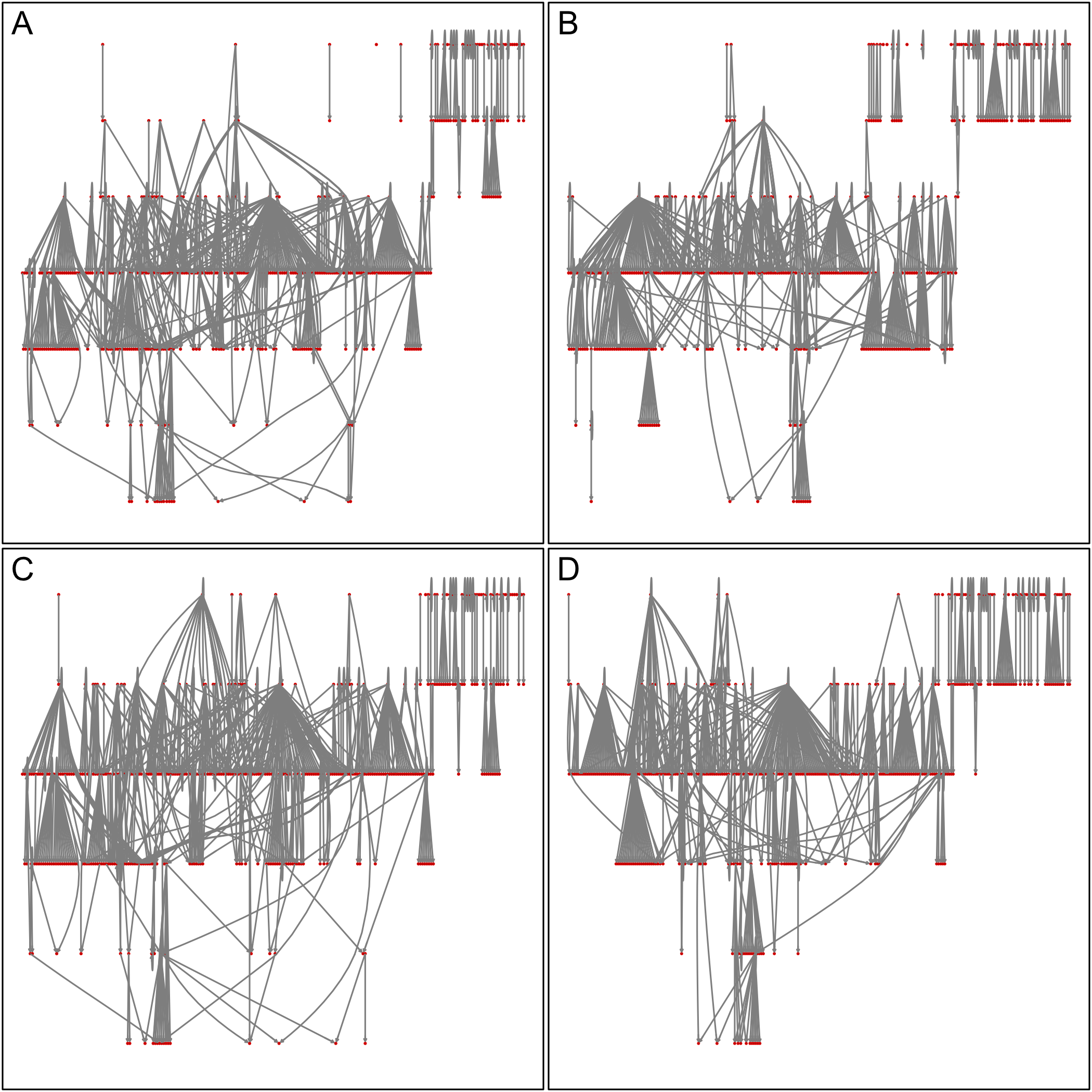} 
\caption{Transcriptional regulation networks generated using informative priors on individual edges (on the basis of {\it E. coli} network).  Red dots:  423 genes  in each network. Prior on individual edges (A), on individual edges and degree distribution (B),  on individual edges and the proportion of feed-back loops; D: all three priors together.} 
\label{FigureInformative}
\end{center}
\end{figure}

If we now turn to networks simulated with an informative prior on individual edges (Figure \ref{FigureInformative}), we see a striking difference with Figure \ref{FigureFlat}. The structure of those networks, even if random, is quite close to the actual {\it E. coli} transcriptional regulation network (Figure \ref{FigurePriorGraph}). The difference between the networks with a prior on degrees (panels B and D) or without (panels A and C) is now more subtle. Actually the prior on individual edges is strong enough to impose a correct distribution of degrees, even if the degree distribution is not specified (see Figure \ref{FigureDeg}). In that Figure, a deviation of the actual number of degrees from the power law, for high degrees, can be observed and is well simulated. A similar behavior of degree distribution can be found in (Dobrin \textit{et al}, 2004). In Figure \ref{FigureInformative}, the frequency of motifs is also controlled directly by the edges probabilities: The number of FBLs is about constant at $1.0 \; 10^7$, for approximately $3 \; 10^{10}$ FFLs, hence a proportion of 0.05\%. The differences between chains are small and all those results have a 5\% CV.
 
\section*{Discussion} 

There are two important applications to the generation of semi-random graphs with known properties: {\it i}. Simulating actual networks for hypothesis testing, software bench-marking, statistical power calculations {\it etc.} (Van den Bulcke, 2006); {\it ii}. Assessing whether a graph is coherent with our prior knowledge in numerical data analytic methods such as Bayesian network modeling, Gaussian graphical methods {\it etc.} Such methods, in their naive implementation, are known to suffer badly from the curse of dimensionality. However, prior knowledge about network structure increases daily, at least for biological networks, and a proper accounting of such knowledge is our only hope to redeem the curse we face.

We have extended here the results presented in Mukherjee and Speed (2008) by including a flexible specification of edge probability, via Bernoulli priors, and by defining a prior on network motifs. In doing so we have dropped the commonly used distance penalty from a pre-specified reference network (whereby "distance" correspond to the number of differing edges between a proposed graph and the reference graph). Such a distance penalty is simple to specify and compute, but has several drawbacks: It is quite coarse and does not give a weight to the various edges, while in fact we may be more or less certain about some of them. Therefore it treats in the same way edges known to be absent or present, and edges for which the we do not know whether they are absent or present. Also, only one reference network is specified and this limits the number of questions we can ask. It is finally more an {\it ad hoc} penalty function than a proper distribution, although that may be seen as a purely technical argument. In any case, an edge by edge prior probability assignment is not much more difficult to specify, it can be simplified by giving default vague probability values if information is lacking, and can be quite powerful when information is available.

There are, however, cases where higher levels of structure are important. In fact, we can hypothesize that higher levels are more important than we usually suspect: That is the whole point of systems biology. Power law degree distribution is a well known characteristic of biological networks. The frequency of occurrence of particular network motifs is another point in case. Fascinating recent work has addressed the question of degree distribution (Leskovec \textit{et al},  2010), but the problem remains for motifs since no direct sampling or generative method is available in that case. We have shown here how to use MCMC sampling to obtain the desired random graphs, even for realistically large number of nodes. An important point is that a sample of random graphs is much more informative than a single approximate, or even "best", estimate graph. With a single graph, all sense of uncertainty is lost, and only over-confidence is gained. Ensemble results are much more robust and useful; but they can be cumbersome and the question is how to best handle them.

Stochastic simulations also give insights about the relative weights of the various components of our prior knowledge. A first point is that prior knowledge about the probable number of edges, or at least about network sparsity, can strongly constrain the set of admissible networks. At least, that is the case when implemented in a form of Bernoulli priors on individual edges. Actually, more flexible priors ({\it e.g.}, hierarchical) could be used instead of Bernoulli to allow more uncertainty about that expected number of edges. Specific knowledge about subsets of high probability, or conversely low probability, edges is also very informative. The degree distribution of a network may appear as a weak predictor of its structure since, for example, with a given number of nodes, two graphs with the same degree distribution may have completely different edge lists. However, we found that specifying a degree distribution has a visible impact on the network structure. Here again it would be easy to be more flexible about that distribution. Imposing a prior on the occurrence of specific motifs can be important in terms of functionality, but leads to more subtle modifications. Note however, that we imposed a distribution on the relative frequency of two loop motifs, rather than on there absolute number. That would be an easy extension, which could have profound consequences on the network structure.
Overall, there are many variables with which the prior can play, and even potential conflicts between components of our prior knowledge, in particular if a strongly informative prior is placed on individual edges. Such conflicts can be hard to figure out without the help of simulations, because our intuition often fails in high dimension and in the case of graphs (Helbing, 2013). In that respect, the possibility to perform quickly billions of iterations for network of sizes commensurable to those of genomes is encouraging. A word of caution is in order here, however: With 423 nodes, a billion simulations correspond to $10^9 \times 423^{-2}$, {\it i.e.} about 6000 full updates of the network. With $10^4$ nodes we might have to go to trillions of simulations to get to convergence and this would currently take us a day and half of computation, although GPU computing, for example, could again increase speed.

To be more precise about the relative weight of the different priors would have required some evaluation of the number of possible graphs, or some form of enumeration of the number of different graphs sampled during stochastic simulations. However there are, for example $2^{178929}$ different directed graphs with 423 nodes, and tracking the list of graph sampled would entailed considerable time and memory capacity.

Obviously, it would be interesting to extend those results to the important problem of statistical learning of the network structure from data acquired in large scale genotyping or phenotyping studies, for example. In a Bayesian context, that simply entails the computation of a data likelihood function, or its marginalisation. However, the models currently favored for that purpose: Gaussian graphical networks (Krumsiek \textit{et al}, 2011; Liu \textit{et al}, 2012) and Bayesian networks (Mukherjee and Speed, 2008), are either undirected or acyclic, respectively. Dealing with undirected graphs would be easy, but the acyclicity of Bayesian networks is quite restrictive. Loops motifs cannot exist formally in such models, unless they are made dynamic, and checking for acyclicity at every iteration imposes significant computational burden. The possibility of using hybrid models (Silva and Ghahramani, 2009) is an interesting possibility to explore.

\section*{Materials and methods}

Graph probabilities such as $P_{Total,G}$ are only defined up to a multiplicative constant. In that case, a simple way to generate sample graphs according to that probability distribution is to use the Metropolis-Hasting sampler (Casella and Robert, 2004). From a current graph $G$, with total probability $P_{Total,G}$ (eq. (\ref{Total-Proba})), a graph ${\tilde G}$ is proposed by first selecting two nodes, say $v_{i}$ and $v_{j}$ of $G$ ($v_{i}$ may be equal to $v_{j}$ in case of auto-loops) and then by deciding on the presence of an edge from $v_{i}$ and $v_{j}$ by a random Bernoulli draw with edge probability $p_{i,j}$. That amounts to sampling ${\tilde G}$ from the Bernoulli prior on edges defined in eq. (\ref{Bernoulli-Proba}). The total probability $P_{Total,\tilde G}$ of ${\tilde G}$ is computed using eq. (\ref{Total-Proba}) and ${\tilde G}$ is accepted with a probability equal to min(1, $P_{Total,\tilde G}$ / $P_{Total,G}$). In case of rejection, ${\tilde G}$ is discarded, and $G$ is again the current sample (that implies that the same graph can be drawn several times in succession). The procedure is iterated as many times as it is needed to reach convergence in probability to the target distribution sought. Convergence can be checked by running several simulation chains and computing Gelman and Rubin's \^{R} criterion (Gelman and Rubin, 1992) on each element of the graphs' adjacency matrix, or by monitoring the degree distributions obtained, or the motifs probabilities in those independent chains.

A C language version of the algorithm has been implemented as a module of the free {\it GNU MCSim} software (Bois, 2009) (http://www.gnu.org/software/mcsim). That software was used for all the simulations presented here. Graphs were produced with R, version 2.14 (R Development Core Team, 2011).

\section*{Acknowledgments }
The research leading to these results has received funding from the scientific council of the Universit\'e de Technologie de Compi\`egne (project Prior-Motives) and the Innovative Medicines Initiative Joint Undertaking, under Grant Agreement number 115439 (StemBANCC), resources of which are composed of financial contribution from the European Union Seventh Framework Programme (FP7/2007-2013) and EFPIA companies in kind contribution. This publication reflects only the author's views and neither the IMI JU nor EFPIA nor the European Commission are liable for any use that may be made of the information contained therein.

\section*{Author Contributions}
F.B. and G.G. designed and performed research, and wrote the paper.

\bibliographystyle{msb}

\end{document}